\documentclass{aa}

\usepackage{graphicx}
\usepackage{natbib}
\usepackage{hyperref}
\usepackage{txfonts}

 \newcommand{\Mo}{$M_\odot$}
 \newcommand{\Ro}{$R_\odot$}
 \newcommand{\arcs}{$^{\prime\prime}$}

\begin{document}

   \title{Unique sextuple system: 65 UMa\thanks{
Tables A1 to A4 are only available in electronic form at the CDS via anonymous ftp to
cdsarc.u-strasbg.fr (130.79.128.5) or via http://cdsweb.u-strasbg.fr/cgi-bin/qcat?J/A+A/}}

   \titlerunning{Unique sextuple system: 65 UMa}

   \author{P. Zasche \inst{1}
            \and R. Uhl\'a\v{r}\inst{2}
            \and M. \v{S}lechta\inst{3}
            \and M. Wolf \inst{1}
            \and P. Harmanec \inst{1}
            \and J.A. Nemravov\'a \inst{1}
            \and D. Kor\v{c}\'akov\'a \inst{1}}

   \authorrunning{P. Zasche et al.}

   \institute{Astronomical Institute, Faculty of Mathematics and Physics, Charles University Prague,
   CZ-180 00 Praha 8, V Hole\v{s}ovi\v{c}k\'ach 2, Czech Republic, \email{zasche@sirrah.troja.mff.cuni.cz}
   \and
   Private Observatory, Poho\v{r}\'{\i} 71, 25401 J\'{\i}lov\'e u Prahy, Czech Republic
   \and
   Astronomical Institute, Academy of Sciences, Fri\v{c}ova 298, CZ-251 65, Ond\v{r}ejov, Czech Republic
   }

   \date{\today}

  \abstract
   {The study of stellar multiple systems provides us with important information about the stellar formation
   processes and can help us to estimate the multiplicity fraction in the Galaxy. 65~UMa belongs to a rather
   small group of stellar systems of higher multiplicity, whose inner and outer orbits are well-known. This
   allows us to study the long-term stability and evolution of the orbits in these systems.}
   {We obtained new photometric and spectroscopic data that when combined with interferometric data
   enables us to analyze the system 65~UMa and determine its basic physical properties.}
   {We perform a combined analysis of the light and radial velocity curves, as well as the period variation
   by studying the times of the minima and the interferometric orbit. A disentangling technique is used
   to perform the spectra decomposition. This combined approach allows us to study the long-term
   period changes in the system for the first time, identifying the period variation due to the motion on
   the visual orbit, in addition to some short-term modulation.}
   {We find that the system contains one more component, hence we tread it as a sextuple hierarchical
   system. The most inner pair of components consists of an eclipsing binary orbiting around a barycenter
   on a circular orbit, both components being almost identical of spectral type about A7. This pair orbits
   on an eccentric orbit around a barycenter, and the third component orbits with a period of about 640~days.
   This motion is reflected in the period variation in the minima times of the eclipsing pair, as well as in
   the radial velocities of the primary, secondary, and tertiary components. Moreover, this system orbits
   around a barycenter with the distant component resolved interferometrically, whose period
   is of about 118~years. Two more distant components (4\arcs and 63\arcs) are also probably gravitationally
   bound to the system. The nodal period of the eclipsing-pair orbit is on the order of only a few centuries,
   which makes this system even more interesting for a future prospective detection of changing the depths of
   minima.}
   {We identify a unique solution of the system 65~UMa, decomposing the individual components and even shifting
   the system to higher multiplicity. The study of this kind of multiple can help us to understand the origin of
   stellar systems. Besides 65~UMa, only another 11 sextuple systems have been studied.
   }


  \maketitle


\section{Introduction}

As members of more complex multiple systems, the eclipsing binaries can provide us important
information about their physical properties, as derived from different methods. This is the case
for 65~UMa, a system whose the close components form an eclipsing binary, and additional components
found to be gravitationally bounded to this pair \citep{2004A&A...424..727P}. Thanks to the
combined analysis, we have been able to derive the radii, masses, and evolutionary statuses of the
close components, in addition to some properties of the distant ones. These systems are still very
rare and mostly lie relatively close to the solar system. Only 39 such systems are known where a
close eclipsing binary is a member of a wide visual binary and we know both orbits, their mutual
inclinations, ratio of periods, etc. For instance, the ratio of periods can tell us something about
the long-term stability of the system. These unique systems are the most suitable for studies of
dynamical effects, the short and long-term evolution of the orbits, etc. (see e.g.
\citealt{1975A&A....42..229S}).

The study of systems of higher multiplicity is still relatively undeveloped yet, and can provide
insight into their formation. Moreover, \cite{2005A&A...439..565G} found that the majority of the
early-type stars are found in multiple systems. Star-forming theories are still based on many ad
hoc assumptions and the physical characteristics of the multiple systems can provide strong
constraints on some of them. These can be e.g. the mass ratios of the inner and outer pairs, the
ratio of periods, and inclinations, see for instance (\citealt{2007prpl.conf..133G} and
\citealt{2008MNRAS.389..925T}). In addition, the multiplicity fraction is one of the most crucial
parameters in theoretical models and nowadays we know of only 20 quintuples, 11 sextuples, and 2
septuple systems \citep{2008MNRAS.389..869E}.

\section{The system 65 UMa}

The multiple system 65~UMa (= WDS J11551+4629) consists of four visible components. The angular
distance between the primary and component D (= HR 4561) is about 63\arcs, while the C component is
at a distance of about 4\arcs. The primary 65~UMa AB = DN~UMa, the brightest member of the system,
was also resolved to be a binary via classical micrometric measurements by
\cite{1908LicOB...5...28A}. Since then, many precise interferometric observations of this close
pair were carried out. Moreover, the primary component was discovered to be a variable
\citep{1979IBVS.1648....1G}. Later, \cite{1982IBVS.2068....1G} found the star to be an eclipsing
binary with the orbital period of about 1.7304~d. Its variability was not discovered earlier owing
to its rather shallow eclipses of only about 0.09~mag. This is caused by the presence of other
components in the aperture and therefore a large fraction of the third light, which reduces the
photometric amplitude of the eclipses. The light curve (hereafter LC) analysis presented by
\cite{1986Ap&SS.125..181G} was slightly indicative of an eccentric orbit. The radial velocity curve
(hereafter RV) was analyzed by \cite{1986PASP...98.1312P}. All of these studies found that the
close binary consists of two very similar stars.

The position angle between the A and B components slowly changes, therefore an orbit of this pair
was derived most recently by \cite{1999AAS..134..545A}, who found a period of about 137~yr and a
semi-major axis about 225~mas. The position angle of the pair has changed since this last paper by
about 40$^\circ$, hence a new analysis is required. Moreover, the combined analysis of the visual
orbit together with the times-of-minima variation due to the movement on this orbit and the
radial-velocity variations can reveal some of the other parameters of the orbit and also of both AB
and C components. The more distant C and D components also belong to this multiple system, but
these show no detectable mutual motion and can be assumed to be motionless. The MSC catalog by
\cite{1997A&AS..124...75T} gives their periods 11 kyr and 591 kyr, respectively.

A distance to 65~UMa was derived from the Hipparcos data. \cite{HIP} gives the value of distance $d
= 246 \pm 108$~pc, while the new reduction of \cite{2007A&A...474..653V} presented the value $d =
212 \pm 30$~pc. The combined spectral type of the AB pair was classified as A3Vn by
\cite{1969AJ.....74..375C}, while the spectrum of the D component is A2p
\citep{1963ApJ...138..118S}, and \cite{2010MNRAS.401.1299J} carried out an analysis of this star.
On the other hand, the spectral classification of the C component has never been performed.
Therefore, the mass of the C component is also poorly constrained and the only information about
this star that we have is a rough estimate of the magnitude difference (see below).

\section{Observations and data reduction}

In total, the target was observed on 82 nights: 29 nights for photometry, and 53 nights for
spectroscopy. The complete $BVRI$ light curves of the eclipsing pair were obtained in 2010 at the
private observatory of one of the authors (RU). However, owing to the relatively high brightness of
the target, only a small 254-mm reflector of moderate focal length was used. This telescope was
unable to separate the two 4\arcs~distant components, therefore the resultant LC was a composite
AB+C light curve. The CCD photometric observations were obtained in standard $B$, $V$, and $R$
filters according to the specification of \cite{Bessell1990}. All of the observations used for the
LC were obtained with the same telescope and instrument setup, and the reduction was also
identical. Furthermore, the complete set of minima times used for the analysis is given in
Appendix, two new minima were measured by Petr Svoboda, Czech Republic.

The CCD spectra were obtained at Ond\v{r}ejov observatory, Czech Republic, using the 2.0-m
telescope equipped with a SITe-005 800 $\times$ 2000 CCD detector. These spectra cover a wavelength
region 626 -- 676 nm. All of them were secured between March 2010 and July 2011 and have a
resolving power $R \sim 12700$. Their S/N values range typically between 100 and 300.

For all of the spectra, the wavelength calibration was done using a ThAr comparison spectra
obtained before and after the stellar spectra itself. The data reduction was performed following
the standard procedures of the data reduction package {\sc IRAF\footnote{{\sc IRAF} is distributed
by the National Optical Astronomy Observatories,   which are operated by the Association of
Universities for Research in Astronomy, Inc., under cooperative agreement with the National Science
Foundation.}}. The flatfields were taken in the beginning and end of each night and their means
were used in the data reduction. After then, the radial velocities were obtained with the program
SPEFO (\citealt{1996A&A...309..521H}, \citealt{1996ASPC..101..187S}), using the zero point
correction by measuring the telluric lines. In total, 55 spectra were obtained in this way.

All available data used for the analysis, the photometry, times of minima, spectroscopy, and the
interferometric measurements are also listed (see the CDS tables).

\begin{table}
 \centering
  \caption{The parameters of the visual (A-B) orbit.}  \label{LongOrbit}
  \begin{tabular}{c c c }
\hline
 Parameter  & Aristidi et al.(1999) & This work \\
 \hline
 $p_{A-B}$ [yr]       & 136.538 $\pm$ 8.4    & 118.209 $\pm$ 0.690 \\
 $a_{A-B}$ [mas]      & 225 $\pm$ 18         & 208.2 $\pm$ 9.7  \\
 $A_{A-B}$ [d]        & --                   & 0.0428 $\pm$ 0.0023 \\
 $T_{{A-B}}$          & 2447140.9 $\pm$ 149.7& 2447516.9 $\pm$ 126.8 \\
 $\Omega_{A-B}$ [deg] & 169.7 $\pm$ 4.6      & 92.1 $\pm$ 4.2 \\
 $\omega_{A-B}$ [deg] & 26.9 $\pm$ 2.1       & 202.7 $\pm$ 1.3 \\
 $i_{A-B}$ [deg]      & 39.7 $\pm$ 1.9       & 38.1 $\pm$ 2.4 \\
 $e_{A-B}$            & 0.531 $\pm$ 0.014    & 0.504 $\pm$ 0.006 \\
 \hline
\end{tabular}
\end{table}

\section{Visual orbit and the period analysis} \label{ChapterVisOrb}

To begin with, we analyzed the visual orbit. The orbital motion influences the apparent period of
the inner eclipsing pair, hence the periodic variation in the times of minima are analyzed
according to the visual orbit parameters.

\begin{figure}
  \centering
  \includegraphics[width=89mm]{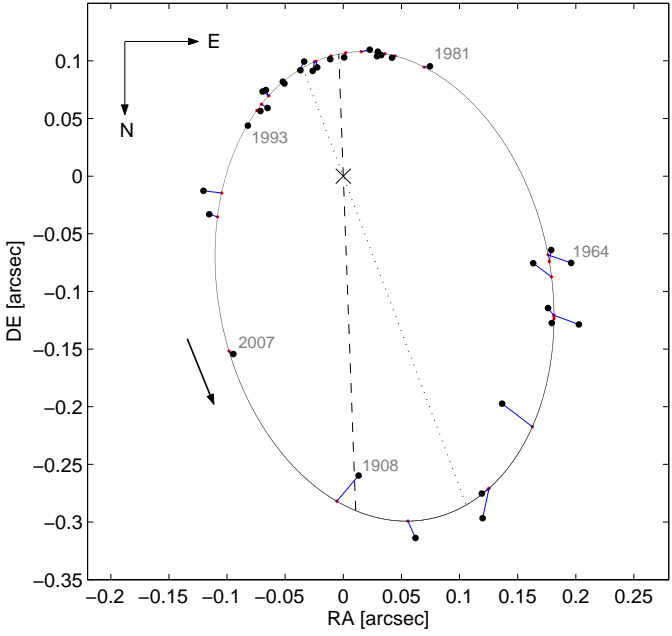}
  \caption{Visual orbit of 65~UMa pair (A-B) as displayed on the sky. The eclipsing binary is
  placed in [0,0]. The dotted line represents the line of the apsides, while the dashed one is the
  line of the nodes. See section \ref{ChapterVisOrb}.}
  \label{FigOrbit}
\end{figure}

Since its discovery as a double, 35 observations of the A-B pair have been obtained. These have
been collected in the Washington Double Star Catalog (hereafter
WDS\footnote{\href{http://ad.usno.navy.mil/wds/}{http://ad.usno.navy.mil/wds/}}, \citealt{WDS}). We
analyzed the data, obtaining the parameters of the visual orbit presented in Table \ref{LongOrbit},
and the final fit together with the data is given in Fig. \ref{FigOrbit}. As one can see from Table
\ref{LongOrbit}, the parameters differ significantly in some aspects. Besides the higher precision,
the most significant difference is found for the orientation of the orbit in space. It is obvious
that the same fit to the data can be obtained with different sets of parameters when we only
interchange the values of two parameters: $(\Omega, \omega) \rightarrow (\Omega + 180^\circ, \omega
+ 180^\circ)$. However, when dealing with astrometric data set only, one cannot distinguish between
these two identical solutions. The only way to do so is to use also the RV data, or the
times-of-minima variation.

For the minimum-times observations, we have only a limited set of data points. If we consider the
period of the A-B pair to be about 118~yr, we have data for only about one-quarter of the orbital
period covered with minima times at the present day. Yet, we can try to carry out an analysis of
these data, by fixing the orbital parameters from astrometry (these in Table \ref{LongOrbit}). We
have regularly observed the minima of this interesting target for the past four years to detect the
period variation.

\begin{table}
 \caption{Final parameters of the short (Aa-Ab) orbit.}
 \label{LITE640d}
 \small
 \centering
 \begin{tabular}{c c c c}
 \hline\hline
 Parameter & Value \\
 \hline
 $p_{Aa-Ab}$ [d]       & 641.5 $\pm$ 16.7 \\
 $A_{Aa-Ab}$  [d]      & 0.00621 $\pm$ 0.00147 \\
 $T_{Aa-Ab}$           & 2449615.4 $\pm$ 38.9 \\
 $\omega_{Aa-Ab}$ [deg]& 0.0 $\pm$ 15.2 \\
 $e_{Aa-Ab}$           & 0.169 $\pm$ 0.048 \\
 \hline
 \end{tabular}
\end{table}

\begin{figure}
  \centering
  \includegraphics[width=89mm]{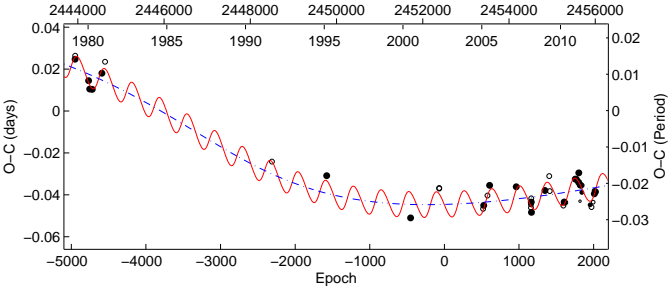}
  \caption{Period variations of the eclipsing pair. Primary minima have been plotted as dots, and
  secondary as circles. The dashed line represents the 118 yr orbit, while the solid one is the
  640 day orbit.}
  \label{FigOC}
\end{figure}

\begin{figure}
  \centering
  \includegraphics[width=89mm]{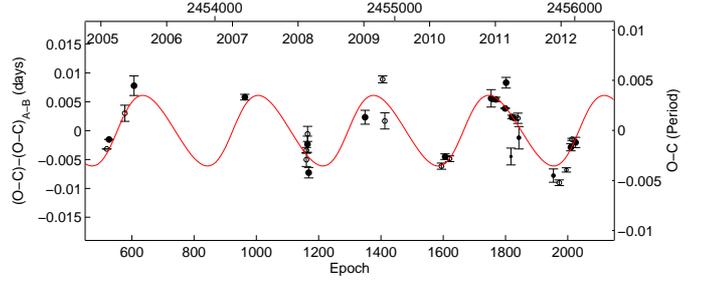}
  \caption{Period variation in the times of minima after subtraction of the 118 yr term. Only the variation
  caused by the component Ab and the most recent minima have been displayed.}
  \label{FigOCdetail}
\end{figure}

This led to an interesting finding that there is also an additional variation. We therefore
analyzed our data set (43 data points in total) assuming two periodic terms. We used the
LIght-Time-Effect hypothesis (hereafter LITE, described e.g. by \citealt{Irwin1959}). The results
of our analysis can clearly be seen in Figs. \ref{FigOC} and \ref{FigOCdetail}. The long-term
periodic modulation (blue) is caused by the 118 yr visual orbit, while the short-period one is the
newly discovered orbit, whose final parameters derived from our analysis are given in Table
\ref{LITE640d}. This variation is clearly visible especially in the more precise recent data points
after subtraction of the long-period term, see Fig. \ref{FigOCdetail}. Here we use the following
labeling of the components: \emph{Aa1} and \emph{Aa2} for the eclipsing binary components,
\emph{Ab} for the 640 day orbit, and \emph{B} for the 118 yr orbit (i.e. following the WDS
notation). Thanks to the high precision of our new observations, the hypothesis of a non-circular
orbit for 65~UMa eclipsing pair was ruled out. The complete list of times of minima together with
the original $BVRI$ photometry are given in electronic-only tables (see Appendix). There is a
problem with some of the minima times, whose accuracy is not always given, hence we cannot perform
a reliable chi-square test of the 640 day hypothesis. However, using the weightening scheme for the
data points, the LITE fit based on the 640 day hypothesis gives the sum of square residuals
0.00318, while disregarding the possibility of a 640 day period the sum is 0.01222.

Using the approach of combining the two LITE terms, one can also derive the parallax of the system
independently of the Hipparcos value and the total mass of the system. The method is as follows:
$(A_{A-B},a_{A-B}) \Rightarrow \pi \Rightarrow M_{\mathrm{tot}}.$ To briefly describe the method,
the amplitude of LITE and the angular semi-major axis of the visual orbit are directly connected
via the parallax \citep{Mayer1990}. Using our new computed parallax and the Kepler's third law, we
calculated the total mass of the system (e.g. \citealt{2001icbs.book.....H}). The values presented
in Table \ref{LITE640d} and the LITE semi-amplitude $A_{A-B}$ were calculated using this approach.
This means that the values of $p_{A-B}, T_{A-B}, \Omega_{A-B}, \omega_{A-B}, i_{A-B}$, and
$e_{A-B}$ were fixed, but the parameters $A_{A-B}, p_{Aa-Ab}, A_{Aa-Ab}, T_{Aa-Ab},
\omega_{Aa-Ab},$ and $e_{Aa-Ab}$ were fitted as free parameters. From this analysis, a new value of
the parallax $\pi = 4.28 \pm 0.49$~mas was obtained, which yields the distance $d = 234 \pm 29$~pc.
Such a value of parallax is slightly lower than the Hipparcos value ($\pi_{\mathrm{Hip}} = 4.72 \pm
0.58$~mas). Using this new value of the parallax, we than computed the total mass of the system
$M_{\mathrm{tot}} = 8.25 \pm 1.85~M_\odot.$ \cite{1999AAS..134..545A} found that $M_{\mathrm{tot}}
= 9.1 \pm 11.6~M_\odot.$

The relative motion of the component C around AB is very slow, but detectable. During more than 200
years of observations, over 60 measurements were obtained (see the WDS) that revealed a change in
position angle of about 5$^\circ$. We analyzed these data, determining a period of longer than
14000~yr. However, this result is very preliminary owing to the poor coverage of only 1/64 of the
orbital period.

\section{Light and radial velocity curves}

To analyze the LC and RV curves, we had to consider a precise ephemeris of the eclipsing pair.
These followed from the minimum times analysis and resulted in the elements for the primary minima
\begin{equation}
  \mathrm{HJD} = 2455651.4491(5) + 1.7304736(32) \cdot \mathrm{E}.
\end{equation}
However, the analysis was not straightforward because of the additional components in the system.
We assumed that the C component does not affect the spectra significantly, since its distance is
about 4\arcs. The observing conditions and seeing were usually better than 2\arcs~during most
nights. Hence, four other components were found to be present in each of the spectra. We used the
program {\sc KOREL} \citep{2004PAICz..92...15H} to disentangle the spectra.

To perform the spectral disentangling, the orbital elements of both orbits were fixed. Hence, the
most crucial for the analysis were the values of the mass ratios and amplitudes of the radial
velocity curves. The ephemerides of the close eclipsing pair were also kept fixed because these are
much more reliably known from the minima-times analysis.

\begin{figure}
  \centering
  \includegraphics[width=89mm]{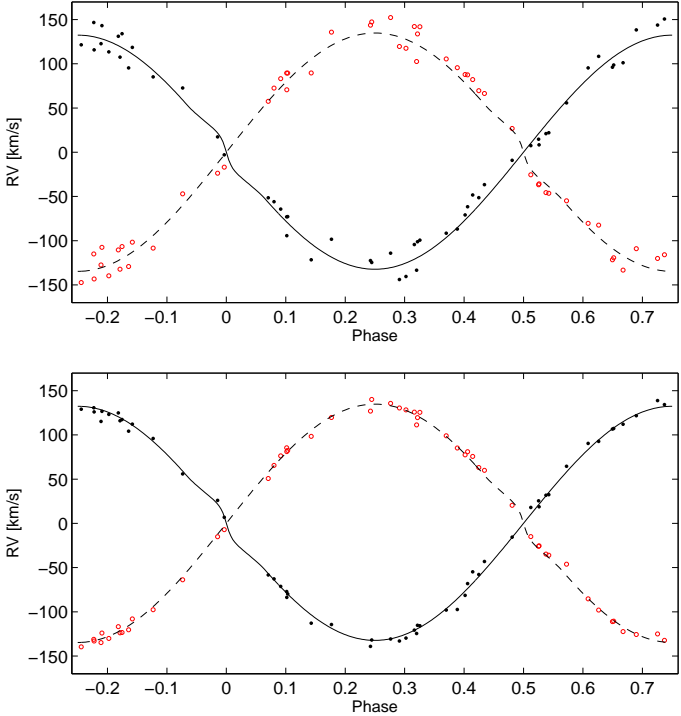}
  \caption{Radial velocity curves of the 65~UMa eclipsing pair. The upper plot shows the original RVs
  before correction for the 640 day variation. The bottom plot shows the RVs of the eclipsing pair
  after the correction.}
  \label{FigRV}
\end{figure}

The parameters $p_{Aa-Ab}$, $T_{Aa-Ab}$, $\omega_{Aa-Ab}$, $e_{Aa-Ab}$, $p_{A-B}$, $T_{A-B}$,
$\omega_{A-B}$, and $e_{A-B}$ were fixed. The results of the other relevant parameters using {\sc
KOREL} are listed in Table \ref{KORELpar}. Despite the results of the mass ratios not being very
conclusive, we were able to make some preliminary estimations of the masses of the individual
components, as described below in Sect. \ref{Discussion}. Since {\sc KOREL} does not provide an
error estimation, the errors in the individual parameters given in Table \ref{KORELpar} resulted
from the following analysis. Several solutions in {\sc KOREL} were calculated, from which only
those with $\chi^2$ value closer than 5\% from our best solution were considered. The errors in the
parameters were assumed to be the maximum difference between these different solutions.

\begin{table}
 \caption{Parameters from the {\sc KOREL} analysis.}
 \label{KORELpar}
 \centering
 \begin{tabular}{c c}
 \hline\hline
 Parameter & Value \\
 \hline
 $q_{Aa1-2}$ ($=\!{M_{Aa2}}/{M_{Aa1}}$) & 0.995 $\pm$ 0.012 \\
 $K_{Aa1}$ [km/s] & 133.3 $\pm$ 4.2 \\
 $K_{Aa2}$ [km/s] & 135.7 $\pm$ 4.2 \\
 $q_{Aa-Ab}$ ($=\!{M_{Ab}}/{M_{Aa1+Aa2}}$) & 0.69 $\pm$ 0.11 \\
 $K_{Ab}$ [km/s] & 19.9 $\pm$ 2.7 \\
 $q_{A-B}$ ($=\!{M_B}/{M_{A}}$) & 0.42 $\pm$ 0.14 \\
 $K_{B}$ [km/s] & 0.41 $\pm$ 0.30 \\
\hline
 \end{tabular}
\end{table}

The program {\sc KOREL} enables us to obtain the RVs of the individual components, which can be
used for some further analysis. We used our knowledge of the ephemerides of the inner pair and the
orbital parameters of the third body (i.e. 640 day orbit), to subtract the 640 day term from the
RVs of the eclipsing pair. This can be clearly seen in Fig. \ref{FigRV}, where the upper plot
represents the original radial velocities, while the bottom plot represents the velocities after
the subtraction of this term. We achieved significant improvement in the quality of the RVs, which
could be used to perform a combined LC and RV analysis.

\begin{figure}
  \centering
  \includegraphics[width=89mm]{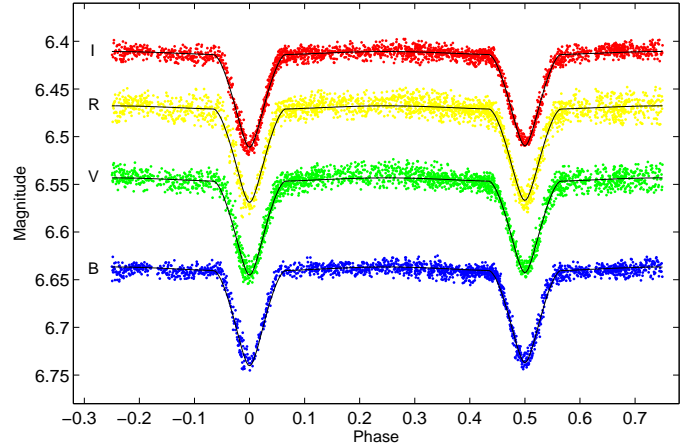}
  \caption{$BVRI$ light curves of 65~UMa together with the final {\sc PHOEBE} fit. The light curve
  parameters are given in Table \ref{LCparam}.}
  \label{FigLC}
\end{figure}

The LC and RV curves of the eclipsing pair were analyzed using the program {\sc PHOEBE}
\citep{Prsa2005}, which is based on the Wilson-Devinney program (WD, \citealt{Wilson1971}). The
derived quantities are given in Table \ref{LCparam}, while the LC has been plotted in
Fig.\ref{FigLC}. The values of synchronicity $F_i$ for both eclipsing components were not derived
from the combined LC and RV analysis. These were calculated from the spectra, and used to compute
the values $v \sin i$ yielding the $F_i$ values. The errors in $F_i$ are the standard deviations in
$F_i$ measured for different spectra and different lines. The limb darkening was approximated using
a linear cosine law, and the values of $x_i$ were interpolated from the tables given in
\cite{vanHamme1993}. In Table \ref{LCparam}, we used the labeling of the two eclipsing components 1
and 2 in the indices instead of Aa1 and Aa2 for a better clarity.

The primary temperature was fixed at value $T_1 = 8000~$K, which agrees with both the A7 spectral
type \citep{2000asqu.book.....C} and the masses of the components (see Table \ref{LCparam}).
Therefore, it is legitimate to ask why a spectral classification of an A3 star was made for the
eclipsing binary \citep{1969AJ.....74..375C}. This is due to presence of the component Ab on a 640
day orbit, whose spectral type is probably of A3 and its light dominates the spectra. Moreover,
\cite{1986Ap&SS.125..181G} speculated that the eclipsing binary components might be of A8-9
spectral type. The large value of the third light is also well-established owing to the three
components that are present in the photometric aperture (in our notation, components Ab+B+C).

\begin{table}
 \caption{Light curve parameters of 65~UMa eclipsing pair.}
 \label{LCparam}
\small
 \centering
 \begin{tabular}{c c}
 \hline\hline
 Parameter & Value \\
 \hline
 $T_1$ [K]   & 8000$^*$ \\
 $T_2$ [K]   & 7948 $\pm$ 20 \\
 $i$  [deg]  & 86.5 $\pm$ 0.2 \\
 $g_1 = g_2$ & 1.00$^*$ \\
 $A_1 = A_2$ & 1.00$^*$ \\
 $F_1$       & 0.423 $\pm$ 0.094 \\
 $F_2$       & 0.384 $\pm$ 0.077 \\
 $L_1$ (B) [\%] & 10.7 $\pm$ 0.4 \\
 $L_2$ (B) [\%] &  9.7 $\pm$ 0.3 \\
 $L_3$ (B) [\%] & 79.6 $\pm$ 1.5 \\
 $L_1$ (V) [\%] & 10.7 $\pm$ 0.5 \\
 $L_2$ (V) [\%] &  9.8 $\pm$ 0.5 \\
 $L_3$ (V) [\%] & 79.5 $\pm$ 2.1 \\
 $L_1$ (R) [\%] & 10.7 $\pm$ 0.7 \\
 $L_2$ (R) [\%] &  9.9 $\pm$ 0.7 \\
 $L_3$ (R) [\%] & 79.5 $\pm$ 3.3 \\
 $L_1$ (I) [\%] & 10.6 $\pm$ 0.4 \\
 $L_2$ (I) [\%] &  9.9 $\pm$ 0.4 \\
 $L_3$ (I) [\%] & 79.5 $\pm$ 1.8 \\ \hline
 \multicolumn{2}{c}{Derived quantities:} \\
 $R_1$ [\Ro] & 1.86 $\pm$ 0.08 \\
 $R_2$ [\Ro] & 1.81 $\pm$ 0.08 \\
 $M_1$ [\Mo] & 1.74 $\pm$ 0.06 \\
 $M_2$ [\Mo] & 1.71 $\pm$ 0.06 \\ \hline \hline
 Note: $^*$ - fixed.
 \end{tabular}
\end{table}

\begin{figure}
  \centering
  \includegraphics[width=89mm]{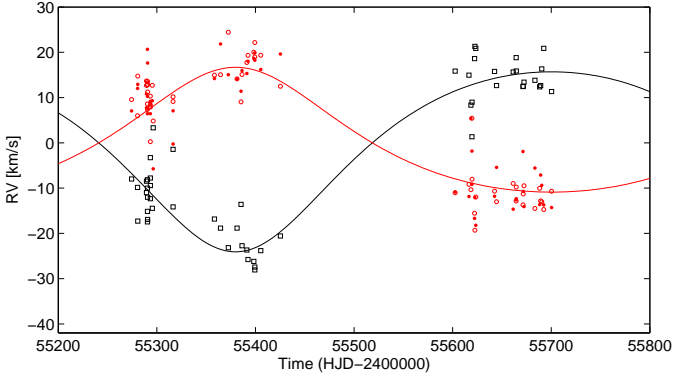}
  \caption{Radial velocity curves on the 640~day orbit. The black squares stand for the Ab
  lines in the spectrum, while the red circles represent the radial velocity residuals after
  subtraction of the eclipsing pair RV curve  (filled for primary, open for secondary).}
  \label{FigRV3}
\end{figure}

The component Ab dominates the spectrum. This body has a well-defined orbit, hence its lines can
also be plotted with the 640 day period. On the other hand, Ab orbits around a barycenter with the
eclipsing pair. We can also plot the residuals from the LC fit from the eclipsing pair (from the
original RVs), which should vary in anti-phase with respect to the Ab lines. This is shown in
Fig.\ref{FigRV3}, where we have plotted the 640 day fit to our spectra, which were acquired over
two consecutive seasons 2010 and 2011.

Moreover, during the photometric monitoring of 65~UMa, two new variables were identified in the
field. One of them was HD~103795 (spectrum K2III, according to \citealt{1962AJ.....67...37U}),
while the other one was SAO~43913 (spectrum F0, according to \citealt{1959AAHam...5..105S}).
Neither was ever reported to be a variable, despite both having been observed by the Hipparcos
satellite. However, our CCD photometry indicates that both are probably variable with amplitudes a
few hundreds of magnitude. SAO~43913 is probably a pulsating star (maybe $\delta$~Sct) with a
period of about three hours, but the type of variability of HD~103795 remains unclear.

\section{Discussion and conclusions} \label{Discussion}

\begin{figure}
  \centering
  \includegraphics[width=89mm]{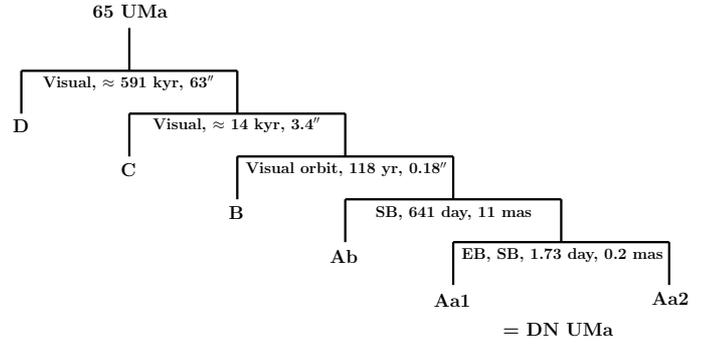}
  \caption{Schematic structure of the whole system 65~UMa.}
  \label{Picture}
\end{figure}

We have performed our first attempt to perform a detailed combined solution of all available data
for 65~UMa, namely photometry, spectroscopy, and interferometry, obtaining quite a reliable picture
of this unusual sextuple hierarchical system (see Fig. \ref{Picture}).

\begin{figure}
  \centering
  \includegraphics[width=89mm]{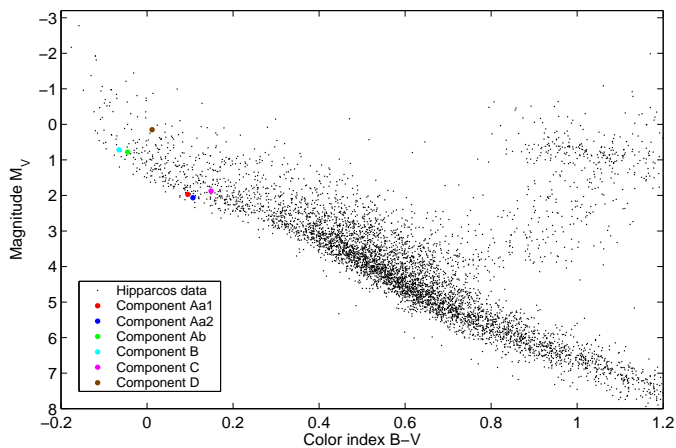}
  \caption{Color-magnitude diagram for all components of the system. Their position is compared with
  the Hipparcos stars (small black dots).}
  \label{Color-mag}
\end{figure}

The inner close eclipsing pair consists of two almost-identical stars of A7 spectral type. This
finding is consistent with the photometric indices $B-V$, $V-R$, and $R-I$ being constant for the
whole phase of the eclipsing binary at a level of 0.005~mag. The stars move on circular orbits with
periods of about 1.73043~d, both being located on the main sequence. Thanks to the combined
analysis, we were also able to compute its distance as $d = 234 \pm 29$~pc, independently of the
Hipparcos satellite data. The 640 day orbit was confirmed by both the minima times and the RV
variations. Applying the spectral disentangling and rough estimation of the mass ratios from this
analysis, one can estimate the masses of the outer components. The Ab component is probably of A1
spectral type and has a mass of about 2.4~$\pm$~0.4~\Mo, from the total mass of the 118 yr visual
orbit, we can estimate the mass of the fourth component (in WDS named B), to be about
2.4~$\pm$~2.0~\Mo. If we assume both these masses of Ab and B components, we can estimate the
magnitude difference. \cite{1999AAS..134..545A} found this value to be 1.9~$\pm$~0.1~mag, while
here we derive 0.7~$\pm$~4.5~mag. The very large error is due to the large uncertainty in the mass
of the fourth body. Another approach is to use the standard mass-luminosity relation and derive the
individual luminosities of the components. Using this approach, we have plotted Fig.
\ref{Color-mag}, where all components of the system 65~UMa are placed in the color-magnitude
diagram. As one can see, the two eclipsing binaries are slightly under-luminous, while the D
component seems to be over-luminous. The same finding about its higher luminosity was found
elsewhere, e.g. \cite{2010MNRAS.401.1299J} or \cite{2007A&A...475.1053A}.

However, some properties of the Aa-Ab orbit remain unclear, such as the inclination angle between
the orbits. We can do some rough estimation of this quantity. The {\sc KOREL} $K_{Aa}$ value and
the predicted amplitude of radial-velocity variations from the LITE$_{Aa-Ab}$ are connected via
$\sin{i_{Aa-Ab}}$. Hence, we obtain $i_{Aa-Ab} \approx 47^\circ,$ which lies well between the
inclinations $i$ of the eclipsing pair and the $i_{A-B}$ of the visual orbit. Nevertheless, its
error is large but this is still only an estimation. We can also compute the predicted minimal
angular separation of the Aa-Ab pair for a prospective interferometric detection. This resulted in
about 11~mas, which is very favorable for modern stellar interferometers, because the magnitude
difference between the Aa and Ab components should also be rather low. On the other hand, the
angular separation of the eclipsing pair components is still rather low, at about only 0.18~mas.


Dealing with a multiple system, we should also consider the nodal period of the close pair and the
640 day orbit, hence the change in the inclination of the eclipsing binary
\citep{1975A&A....42..229S}. The most crucial here is the ratio of periods ${p_{Aa-Ab}}^2/P$, which
implies that the nodal period was about 650~years, a duration that should be practical to observe.
Unfortunately, we do not have a complete set of orbital parameters of the Aa-Ab pair, so this is
only first rough estimation. However, this nodal period is not too long and potentially detectable.
Further observations would help us to detect the change in the eclipse depths. Despite these being
rather shallow, this effect was detected in only nine other cases, hence it would be interesting to
reattempt detections, especially with the modern ultra-precise satellite photometry.

Nevertheless, 65~UMa is a rather unusual system, we presently know of only 11 other sextuple
systems (see \citealt{2008MNRAS.389..869E}). The mass ratio of close to unity for the inner pair
seems to agree with some theoretical models of star formation, e.g. \cite{2003MNRAS.342..926D}.
Moreover, some studies (e.g. \citealt{2002A&A...382..118T}) indicate that about one-third of all
multiples are higher-order systems. \cite{2007prpl.conf..133G} discussed a finding that there is a
difference between the number of observed and expected higher-order multiples (quadruples and
higher). Perhaps the discovery of other systems similar to 65~UMa would diminish this discrepancy.

\begin{acknowledgements}
Dr. Pavel Mayer is acknowledged for a useful discussion and valuable advices. We would like to
thank Mr. P. Chadima for obtaining the two spectra of 65~UMa at Ond\v{r}ejov observatory and P.
Svoboda for sending us his photometric data for 65~UMa. This work was supported by the Czech
Science Foundation grant no. P209/10/0715, by the research programme MSM0021620860 of the Czech
Ministry of Education, and by the grant UNCE 12 of the Charles University in Prague. This research
has made use of the Washington Double Star Catalog maintained at the U.S. Naval Observatory, the
SIMBAD database, operated at CDS, Strasbourg, France, and of NASA's Astrophysics Data System
Bibliographic Services.

\end{acknowledgements}

\Online
\begin{appendix}
\section{Data sets used for the analysis}\label{apendix}
\subsection{Minima times}

\begin{table}[h]
 \caption{Heliocentric minima times used for the analysis.}
  \centering
 \begin{tabular}{l c c c l}
 \hline\hline
 HJD-2400000 & Error & Type & Filter & Reference \\
 \hline
 43936.4851 &       &  Prim & B    & \cite{1982IBVS.2068....1G} \\
 43937.352  &       &  Sec  & B    & \cite{1982IBVS.2068....1G} \\
 44249.683  &       &  Prim & B    & \cite{1982IBVS.2068....1G} \\
 44275.6355 &       &  Prim & B    & \cite{1982IBVS.2068....1G} \\
 44334.4700 &       &  Prim & B    & \cite{1982IBVS.2068....1G} \\
 44557.7034 &       &  Prim & B    & \cite{1982IBVS.2068....1G} \\
 44629.5218 &       &  Sec  & B    & \cite{1982IBVS.2068....1G} \\
 48500.4500 &       &  Sec  & Hp   & \cite{HIP} \\
 49771.44535&       &  Prim & B    & \cite{1997IAPPP..67...30D}  \\
 51716.4305 & 0.0004&  Prim & BV   & \cite{2003IBVS.5476....1B} \\
 52381.7955 & 0.0003&  Sec  & V    & \cite{2009AJ....138..664Z} \\
 52381.7958 & 0.0003&  Sec  & B    & \cite{2009AJ....138..664Z} \\
 53399.2797 &       &  Sec  & V    & \cite{VSOLJ} \\
 53412.2596 &       &  Prim & V    & \cite{VSOLJ} \\
 53499.6511 & 0.0014&  Sec  & C    & \cite{2006IBVS.5690....1K}    \\
 53550.7037 & 0.0017&  Prim & C    & \cite{2006IBVS.5690....1K}    \\
 54166.7367 & 0.0005&  Prim & C    & \cite{2007IBVS.5806....1K}    \\
 54508.4870 & 0.0012&  Sec  & R    & \cite{2009AJ....138..664Z}\\
 54508.4886 & 0.0005&  Sec  & BV   & \cite{2010IBVS.5933....1L}   \\
 54514.5462 & 0.0014&  Prim & BV   & \cite{2010IBVS.5933....1L}   \\
 54515.4132 & 0.0013&  Sec  & V    & \cite{2010IBVS.5933....1L}   \\
 54521.4630 & 0.0009&  Prim & R    & \cite{2009AJ....138..664Z} \\
 54834.68161& 0.0012&  Prim & R    & \cite{2009AJ....138..664Z} \\
 54932.45785& 0.0006&  Sec  & R    & P.Svoboda - this paper  \\
 54944.5637 & 0.0014&  Sec  & V    & \cite{2010IBVS.5933....1L}         \\
 55259.49539&0.00054&  Sec  & BVR  & R.Uhla\v{r} - this paper    \\
 55279.39706&0.00048&  Prim & R    & P.Svoboda - this paper  \\
 55304.48805&0.00050&  Sec  & BVRI & R.Uhla\v{r} - this paper    \\
 55535.51187&0.00151&  Prim & BVI  & R.Uhla\v{r} - this paper    \\
 55561.46825&0.00038&  Prim & VRI  & R.Uhla\v{r} - this paper    \\
 55613.37982&0.0001 &  Prim & R    & \cite{2011OEJV..137....1B}  \\
 55618.57558&0.00092&  Prim & C    & R.Uhla\v{r} - this paper  \\
 55645.38457&0.00146&  Sec  & C    & R.Uhla\v{r} - this paper  \\
 55651.44796&0.00030&  Prim & I    & R.Uhla\v{r} - this paper  \\
 55664.42595&0.00022&  Sec  & I    & R.Uhla\v{r} - this paper  \\
 55683.46083&0.00089&  Sec  & I    & R.Uhla\v{r} - this paper  \\
 55689.51398&0.00194&  Prim & I    & R.Uhla\v{r} - this paper  \\
 55881.58599&0.00116&  Prim & C    & R.Uhla\v{r} - this paper  \\
 55913.59785&0.00049&  Sec  & C    & R.Uhla\v{r} - this paper  \\
 55953.40014&0.00037&  Sec  & R    & R.Uhla\v{r} - this paper  \\
 55978.49541&0.00061&  Prim & V    & R.Uhla\v{r} - this paper  \\
 55984.55331&0.00019&  Sec  & I    & R.Uhla\v{r} - this paper  \\
 56004.45281&0.00089&  Prim & C    & R.Uhla\v{r} - this paper  \\
 \hline
 \end{tabular}
 \end{table}

\newpage

\subsection{Photometry}

\newpage

\subsection{Spectroscopy}

\begin{table}[h]
 \caption{Radial velocities used for the analysis.}
  \centering
 \begin{tabular}{c r r r r r}
 \hline\hline
 HJD-2400000 & $RV_{1,orig}$ & $RV_{2,orig}$ & $RV_3$ & $RV_{1,corr}$ & $RV_{2,corr}$ \\
 \hline
   55700.4062 & -143.91 &  119.58 &  11.32 & -133.04 &  130.45 \\
   55692.4050 &  101.22 & -133.27 &  20.87 &  112.05 & -122.44 \\
   55689.3063 &   85.21 & -108.39 &  12.67 &   96.02 &  -97.59 \\
   55690.4499 &   21.06 &  -45.60 &  16.34 &   31.88 &  -34.78 \\
   55688.3121 & -140.57 &  117.50 &  12.38 & -129.78 &  128.30 \\
   55683.5066 &   14.74 &  -36.74 &  13.83 &   25.48 &  -26.00 \\
   55672.3907 &  -94.32 &   70.58 &  13.40 &  -83.76 &   81.15 \\
   55671.3706 &    7.48 &  -25.50 &  12.43 &   18.02 &  -14.96 \\
   55671.6085 &   96.10 & -121.61 &  12.50 &  106.65 & -111.06 \\
   55661.4470 &  115.79 & -143.33 &  15.66 &  126.10 & -133.01 \\
   55664.4727 &    8.33 &  -35.73 &  18.82 &   18.72 &  -25.34 \\
   55664.5012 &   22.05 &  -46.35 &  15.87 &   32.44 &  -35.96 \\
   55644.5210 &   -2.91 &  -16.85 &  12.67 &    6.87 &   -7.06 \\
   55642.4545 &  113.53 & -139.80 &  15.78 &  123.24 & -130.09 \\
   55623.4771 &   95.38 & -129.19 &  20.87 &  104.25 & -120.31 \\
   55622.2782 & -121.65 &   89.66 &  18.61 & -112.84 &   98.47 \\
   55622.5849 & -133.41 &  102.54 &  21.31 & -124.58 &  111.37 \\
   55619.5604 &   55.86 &  -54.86 &   1.34 &   64.53 &  -46.19 \\
   55619.6977 &   98.67 & -119.32 &   8.98 &  107.35 & -110.65 \\
   55618.5445 &   17.42 &  -23.71 &   8.33 &   26.04 &  -15.10 \\
   55616.5299 &  107.53 & -132.28 &  14.95 &  116.03 & -123.78 \\
   55602.5732 &  121.49 & -147.33 &  15.85 &  129.16 & -139.66 \\
   55425.3137 & -101.14 &  133.75 & -20.57 & -115.30 &  119.58 \\
   55405.3360 &  146.75 & -115.08 & -23.82 &  130.75 & -131.08 \\
   55399.3473 & -104.42 &  142.16 & -28.06 & -120.75 &  125.82 \\
   55399.3633 &  -99.44 &  141.93 & -27.44 & -115.77 &  125.59 \\
   55398.3451 &  150.67 & -115.91 & -26.18 &  134.29 & -132.29 \\
   55392.3562 & -114.19 &  152.26 & -25.77 & -130.79 &  135.67 \\
   55391.3411 &  138.33 & -109.01 & -23.67 &  121.71 & -125.62 \\
   55386.3829 &  134.08 & -106.71 & -22.72 &  117.40 & -123.39 \\
   55385.3757 & -122.49 &  143.68 & -13.58 & -139.18 &  126.99 \\
   55381.3681 &   72.64 &  -47.05 & -18.85 &   55.96 &  -63.73 \\
   55372.4806 &  143.17 & -107.56 & -23.15 &  126.71 & -124.02 \\
   55364.4965 &  -98.38 &  135.79 & -18.84 & -114.43 &  119.74 \\
   55358.3523 &  108.38 &  -82.47 & -16.83 &   92.76 &  -98.08 \\
   55316.4097 &  -86.88 &   95.64 &  -1.42 &  -97.36 &   85.17 \\
   55316.4322 &  -70.94 &   87.93 & -14.14 &  -81.43 &   77.45 \\
   55296.3370 &  122.71 & -127.49 &   3.33 &  115.34 & -134.86 \\
   55295.3948 & -124.67 &  147.41 & -14.47 & -131.89 &  140.18 \\
   55293.3630 &  -51.44 &   57.58 &  -3.28 &  -58.35 &   50.66 \\
   55293.3797 &  -55.91 &   72.52 &  -9.43 &  -62.82 &   65.61 \\
   55293.3993 &  -64.32 &   83.21 & -12.32 &  -71.23 &   76.30 \\
   55293.4198 &  -72.79 &   89.54 &  -7.76 &  -79.71 &   82.62 \\
   55290.4201 &  -91.57 &  105.50 &  -8.15 &  -98.02 &   99.04 \\
   55290.4826 &  -61.63 &   87.50 & -15.13 &  -68.09 &   81.04 \\
   55290.4973 &  -48.36 &   82.18 & -17.47 &  -54.83 &   75.71 \\
   55290.5149 &  -51.31 &   69.63 & -10.02 &  -57.78 &   63.16 \\
   55290.5316 &  -36.63 &   66.50 & -16.93 &  -43.10 &   60.03 \\
   55290.6121 &   -9.17 &   26.90 & -12.01 &  -15.65 &   20.41 \\
   55289.4655 &  131.18 & -110.46 & -10.99 &  124.88 & -116.76 \\
   55289.5066 &  118.55 & -101.70 &  -8.47 &  112.24 & -108.01 \\
   55280.4506 &   95.34 &  -80.39 &  -9.83 &   90.42 &  -85.30 \\
   55280.6528 &  143.80 & -120.04 & -17.29 &  138.85 & -124.99 \\
   55274.3815 &  -73.08 &   89.51 &  -7.98 &  -77.08 &   85.51 \\
  \hline
  \end{tabular}
  \end{table}

\newpage

\subsection{Interferometry}

\begin{table}[h]
 \caption{Positional measurements of A-B pair used for the analysis.}
  \centering
 \begin{tabular}{l l r l}
 \hline\hline
 Year  & $\rho$ [arcs] & $\phi$ [deg] & Reference \\
 \hline
1908.31   &  0.26   &    2.9   & \cite{1908LicOB...5...28A} \\
1919.63   &  0.32   &   11.2   & \cite{1932QB821.A43......} \\
1934.25   &  0.30   &   23.4   & \cite{1961ApJS....6....1K} \\
1934.43   &  0.32   &   22.0   & \cite{1937LicOB..18...53A} \\
1945.36   &  0.24   &   34.7   & \cite{Biesbroeck1954} \\
1958.09   &  0.22   &   54.6   & \cite{vandenBos1960} \\
1958.38   &  0.21   &   57.0   & \cite{vandenBos1960} \\
1958.42   &  0.24   &   57.6   & \cite{vandenBos1960} \\
1962.14   &  0.18   &   65.2   & \cite{1962AJ.....67..555V} \\
1963.546  &  0.19   &   70.3   & \cite{1971PUSNO..22....1W} \\
1964.150  &  0.21   &   69.0   & \cite{1971PUSNO..22....1W} \\
1981.4619 &  0.121  &  142.0   & \cite{1984ApJS...54..251M} \\
1983.4141 &  0.111  &  157.8   & \cite{1987AJ.....93..688M} \\
1984.0530 &  0.110  &  162.7   & \cite{1987AJ.....93..688M} \\
1984.3751 &  0.112  &  164.6   & \cite{1987AJ.....93..688M} \\
1984.3834 &  0.108  &  164.4   & \cite{1987AJ.....93..688M} \\
1985.4894 &  0.112  &  168.3   & \cite{1987AJ.....93..688M} \\
1986.4066 &  0.103  &  179.6   & \cite{1989AJ.....97..510M} \\
1987.2695 &  0.102  &  186.3   & \cite{1989AJ.....97..510M} \\
1988.1653 &  0.097  &  193.4   & \cite{1993AJ....106.1639M} \\
1988.2523 &  0.095  &  196.1   & \cite{1989AJ.....97..510M} \\
1989.1561 &  0.105  &  198.7   & \cite{1997AJ....114.1623F} \\
1989.2272 &  0.099  &  201.9   & \cite{1990AJ.....99..965M} \\
1990.2701 &  0.095  &  212.2   & \cite{1992AJ....104..810H} \\
1990.2756 &  0.097  &  212.4   & \cite{1992AJ....104..810H} \\
1991.3213 &  0.101  &  223.4   & \cite{1994AJ....108.2299H} \\
1991.3269 &  0.100  &  221.7   & \cite{1994AJ....108.2299H} \\
1991.9053 &  0.088  &  227.8   & \cite{1994AJ....108.2299H} \\
1992.3071 &  0.091  &  231.6   & \cite{1994AJ....108.2299H} \\
1993.1970 &  0.093  &  241.9   & \cite{1994AJ....108.2299H} \\
1997.074  &  0.121  &  276.0   & \cite{1999AAS..134..545A} \\
1998.430  &  0.120  &  286.0   & \cite{2001AA...367..865P} \\
2007.3114 &  0.181  &  328.5   & \cite{2009AJ....138..813H} \\
 \hline
 \end{tabular}
 \end{table}

\end{appendix}

\end{document}